\documentstyle[prb,aps,multicol,epsfig]{revtex} 
\begin{document}
  \baselineskip 12 pt \parskip 3 pt
\begin{center}
  {\Large\bf Comparison of History Effects in Magnetization in Weakly pinned\\[.1cm] Crystals of high-$T_c$ and low-T$_c$ Superconductors}\\[.5cm]
      D. Pal, S. Sarkar, A. Tulapurkar, S. Ramakrishnan and A. K. Grover\\
  {\small~Tata Institute of Fundamental Research, Mumbai-400005, India}\\[.2cm]

      G. Ravikumar\\
  {\small~TPPED, Bhabha Atomic Research Centre, Mumbai-400085, India}\\[.2cm]

      D. Dasgupta and Bimal K. Sarma\\
  {\small~Department of Physics, University of Wisconsin, Milwaukee, WI-53201, U.S.A.}\\[.2cm]

      C. V. Tomy, \\
{\small~Department of Physics, Indian Institute of Technology,  Bombay, Mumbai-400076, India}\\[.2cm]

G. Balakrishnan and D. Mck Paul\\
  {\small~Department of Physics, University of Warwick, Coventry, CV4 7AL, United Kingdom}

\end{center} \vspace{.0cm}


{\bf{\bf\it Abstract} --- A comparison of the history effects in weakly pinned single crystals of a high $T_c$ YBa$_2$Cu$_3$O$_{7 - \delta}$ (for H $\parallel$ c) and a low $T_c$ Ca$_3$Rh$_4$Sn$_{13}$, which show anomalous variations in critical current density $J_c(H)$ are presented via tracings
of the minor magnetization hysteresis loops using a vibrating sample magnetometer. The sample histories focussed are, (i) the field cooled (FC), (ii) the zero field cooled (ZFC) and (iii) an isothermal reversal of field from the normal state. An understanding of the results in terms of the modulation in the plastic deformation of the elastic vortex solid and supercooling across order-disorder transition is sought.
}
   \vspace{.25cm}

The critical current density $J_c(H)$ of a hard superconductor does not display any path dependence in field ( H ) and it decays monotonically with increase in H. Its magnetic response   is well described by the prescriptions of the celebrated critical state model. An isothermal magnetization hysteresis (M-H) loop of a strongly pinned superconductor defines an envelope within which lie magnetization values measured along all paths. However, an anomalous magnetization maximum in $J_c(H)$, which relates either to the second magnetization peak (SMP) and/or to the peak effect (PE) phenomenon, occurs ubiquitously in weakly pinned samples of
low $T_c$ and high $T_c$ superconductors. Ever since $J_c(H)$ data were looked at carefully in a clean single crystal of Nb [1], it became evident that $J_c(H)$ depends on the history of  a given H. Early transport data of Steingart et al. [1] in Nb revealed that,
\begin{equation}
$$J_c^{FC}(H)~\ge~J_c^{rev}(H)~\ge~J_c^{ZFC}(H);~H~\le~H_m,$$
\end{equation}
where $H_m$ is the field at which $J_c(H)$ peaks and the rest of symbols have their usual meaning. No exceptions have so far been reported as regards the above inequality in the context of low $T_c$ superconductors which display either the SMP or the PE phenomenon. The collective pinning description due to Larkin Ovchinnikov (LO) [2], which relates $J_c$ inversely to the volume of the collectively pinned Larkin domain $V_c$ ($J_c$~$\propto$~$V_c^{-1/2}$), suffices to rationalize the
observed behavior via the possibility of supercooling the disorder existing at $H_m$ during the field cooled (FC) mode. On the other hand, in the ZFC mode, bundles of vortices invade the superconductor at high velocity and they manage to explore the more ordered configurations (as compared to that in the FC mode). However, this appealing scenario undergoes a phase reversal in samples of high $T_c$ cuprates, where the FC configurations are found to be more ordered than those obtained in the ZFC manner [3]. The inequality statement of eqn. (1) therefore becomes infructuruous. Another circumstance, where this inequality experiences limitations even in low $T_c$ superconductors is where the SMP and the PE are present in the same isothermal scan [4]. In such a case, the path dependence in $J_c(H)$ could be evident beyond the peak position of the SMP, upto the peak position of the PE.

A convenient way to explore path dependence in $J_c(H)$ is the
study of characteristic features of the minor hysteresis loops
(MHL). Existence of linear relationship between $J_c(H)$ and
$\Delta$ M(H) [= M($H^{rev}$) - M($H^{for}$)] facilitates this
task. We shall provide here a glimpse into the results of a
detailed study of three types of MHL in a weakly pinned crystal [3] of YBa$_2$Cu$_3$O$_{7 - \delta}$ (YBCO) for H $\parallel$ c. We shall highlight the differences in the observed behavior with those reported earlier in the low $T_c$ systems [5-7]. The eqn. (1) has been exemplified via study of MHL in samples of 2H-NbSe$_2$ [5], CeRu$_2$ [6] and Ca$_3$Rh$_4$Sn$_{13}$ [7], where only the PE is visible.We shall also present here newer results in a single crystal of Ca$_3$Rh$_4$Sn$_{13}$ (CaRhSn) at a temperature, where both the SMP and the PE are present [4].

In YBCO, MHL were initiated from (i) FC magnetization values
(type-I), (ii) ZFC magnetization values, i.e., M(H$^{for}$) (type-II) and (iii) the magnetization values on the return leg of the M-H loop, i.e., M(H$^{rev}$) (type-III). The insets and the main panels of Figs. 1 and 2 show representative behavior at temperature(s), where anomaly in $J_c(H)$ is broad. In Fig. 1, note that the FC minor curves (type I) in YBCO remain confined inside the envelope loop. This is in contrast to the behaviour in low $T_c$ systems, where the analogous MHL overshoot and cut across the envelope loop exemplifying the inequality, $J_c^{FC}$(H)~$\ge$~$J_c^{ZFC}$(H). In the main panel of Fig.2, the MHL initiated from the forward leg (type II) remain well inside the reverse leg of the envelope loop. On the other hand, the inset panel of Fig. 2 shows that a MHL initiated from the reverse leg (type III) overshoots the forward
envelope loop, thereby exemplifying the inequality,
$J_c$(H$^{rev}$)~$\ge$~$J_c$(H$^{for}$), which is not different from the behaviour well documented in the low $T_c$ systems. Two significant points to note however are as follows : 1. While the MHL of the type I (cf. Fig. 1) initiated from 2 Tesla and 10 Tesla readily merge into the reverse envelope loop, those initiated from fields in between 3 Tesla and 9 Tesla merge into the envelope loop after a field change of about 2 Tesla. This implies that not only the FC configurations have $J_c$ values lower than their
counterparts along the reverse envelope loop, but, also that the metastability effects in $J_c$(H) persist upto 9 Tesla, a field value greater than the notional peak field of the broad SMP at 28.7 K. 2. While the minor loop of the type II initiated from a field of 11 Tesla (see Fig. 2) readily merges into the reverse envelope loop, those initiated from 2 Tesla~$\leq$~H~$\leq$~10 Tesla undershoot the envelope loop, thereby confirming the presence of the metastability effect in $J_c(H)$ upto 10 Tesla.

In a crystal of CaRhSn [$T_c(0)$~$\approx$8.2 K], a SMP can be seen distinct from the PE at T~$\le$~4 K. Fig. 3 shows a plot of the M-H loop in this sample at 1.7 K along with the FC minor loops initiated from fields ranging from far below the onset field of SMP ($H_{smp}^{on}$) to above the maximum of the PE ($H_{pe}^m$). The minor curves initiated from 
1.0~Tesla~$\le$~H~$\le$~2.6 Tesla overshoot the reverse envelope loop, thereby implying that $J_c^{FC}(H)$~$\ge$~$J_c^{rev}(H)$. Minor curves initiated from H~$\ge$~$H_{pe}^m$ at 1.7 K readily merge into the envelope loop, whereas those initiated from H~$\le$~0.6 Tesla undershoot the envelope loop. The former indicates that the path dependence in $J_c(H)$ presumably ceases at $H_{pe}^m$. The latter observation could imply that $J_c^{FC}(H)$~$\le$~$J_c^{ZFC}(H)$ at H~$\le$~0.6~Tesla, a behavior analogous to that seen in YBCO for all fields.
Considering that a FC configuration attempts to freeze in the
disorder persisting at the peak position of the PE (at a given H), it is
pertinent to recall here that an earlier study [5] in CaRhSn showed that the PE surfaces in a robust manner in temperature dependent ac susceptibility runs only for vortex states prepared in H~$\ge$~0.5 Tesla. For 0.35~Tesla~$\le$~H~$\le$~0.5 Tesla, the PE is so nascent that the shrinkage in correlation volume $V_c$ across PE is miniscule, and for H~$<$~0.35 Tesla, no signature of the PE is evident in $\chi^{\prime}_{ac}(T)$ data. This in turn implies that while cooling the given CaRhSn crystal across the superconducting transition in H~$\le$~0.5 Tesla, one would not encounter the pinned amorphous state, which could get supercooled on further lowering the temperature.

Following Kokkaliaris et al. [8], we now examine in YBCO and CaRhSn the progressive change in the plastic deformation of the elastic vortex solid by comparing the saturated value at an appropriate field of a given MHL (of the type II) with the magnetization value corresponding to the same field but lying on a neighboring (i.e., successive) MHL. In Fig. 2, we have identified one such difference, viz., $\Delta$M$_{suc}$ for H = 6 Tesla. Non-zero value of $\Delta$M$_{suc}$ implies progressive enhancement in plasticity. In Fig. 4, we show plots of $\Delta$M$_{suc}$ vs H in YBCO at 26 K and in CaRhSn at 1.7 K. In YBCO, the dislocations causing plastic deformation start to proliferate near the onset position of the anomalous increase in $J_c(H)$ and the plasticity reaches the saturation limit near 9 Tesla (see Fig. 4(a)), above which the metastability effects in $J_c(H)$ cease.

In CaRhSn, $\Delta$M$_{suc}$ vs H (see Fig. 4(b)) shows an interesting modulation with two maxima lying in between the onset and peak positions of the SMP and near the onset position of the PE, respectively. $\Delta$M$_{suc}$ vanishes just above $H_{pe}^m$, where the metastability effects in $J_c(H)$ also cease. The relative heights of the two maxima imply that the vortex matter, after the occurrence of the SMP and near the onset of the PE, is sufficiently well ordered. It is to be noted that $J_c(H)$ monotonically decreases between $H_{smp}^m$ and $H_{pe}^{on}$, thereby indicating that the topologically defective vortex solid existing at $H_{smp}^m$ could heal to some extent while approaching $H_{pe}^{on}$. In this context it could be instructive to view the plot (see Fig. 5) of the difference between the saturated value of a FC minor curve (M$^{sat}_{FC}$) and the corresponding magnetization value (M$^{env}$) on the usual envelope curve as a function of H. One such difference at a field of 2 Tesla has 
been identified in Fig. 3. Fig.5 displays the plot of the parameter R$_{FC}$ [= ($M^{sat}_{fc} - M^{env}$)/$\Delta$M(H)] versus H [8]. Since J$^{FC}_c(H)$ has correlation with current density of the pinned amorphous state existing at the maximum position of the PE, the relative values of this parameter signify how far is a given ZFC vortex state from its FC counterpart. From Fig. 5, it is apparent that the ZFC vortex state at the onset field (H$^{on}_{pe}$) of the PE is relatively more ordered than that at the maximum field (H$_{smp}^m$) of the SMP. The equivalence in the values of the parameter $R_{FC}$ near $H_{smp}^{on}$and $H_{pe}^{on}$ reflects similarity in state of spatial order before the commencement of two anomalous variations in $J_c$. The modulation in $R_{FC}$ versus H reflects the modulation in the plasticity of the vortex solid. Note that the deformed vortex solid is not only far from fully amorphous state at H$_{smp}^m$ but it also heals further between H$_{smp}^m$ and H$_{pe}^{on}$. In addition, the negative values of $R_{FC}$ for H $\leq$ 0.6 Tesla could imply that $J^{FC}_c$~$\leq$~$J^{ZFC}_c$ for these fields.

As yet it is not completely obvious why the FC states in YBCO show behavior different from that in low $T_c$ superconductors. One plausible reason could be a clear distinction and a large separation between the vortex melting and the irreversibility line in the high $T_c$ samples, where the vortices encounter the vortex liquid line first and they enter into a pinned configuration only on crossing the irreversibility line. On the other hand, in low $T_c$ systems, the vortex state at $H_{pe}^{m}$ is termed as pinned amorphous and this amorphous state loses its pinning characteristic at the irreversibility field, which lies infinitesimally close to the $H_{c2}$ line, where the vortices get nucleated.
An understanding of the history effects evident via tracings of MHL in low T$_c$ superconductors was provided by G. Ravikumar \textit{et al.} [5] through a phenomenological model incorporating path dependence in J$_c(H)$ via the relationship, $J_c(B+\Delta~B) = J_c(B) + (|\Delta~B|/B_r)(J_c^{st}-J_c)$. In the limit that the parameter B$_r$ measuring the extent of metastability vanishes, the above model reduces to the usual critical state model. A key assumption of this model is the postulation of stable vortex state with current density J$_c^{st}$, which can be reached by repeated cyclings of the field by an amount $\Delta$B. Prior to the PE the ZFC vortex states are the stable ones, whereas in between H$_{pe}^{on}$ and H$_{pe}^m$, the inequality $J_c^{ZFC}~<~J_c^{st}~<~J_c^{rev}$ should hold. An elucidation of this fact via tracing of stable hysteresis loop above the PE region in weakly pinned crystals of 2H-NbSe$_2$ and CeRu$_2$ has been recently demonstrated [6,9].
\linebreak
\linebreak
\centerline{REFERENCES}
\begin{small}
\vspace{-0.2 cm} 

\end{small}
\begin{figure} 
\caption{Portions of the minor hysteresis loops (dotted curves)  initiated from the FC magnetization values along with the envelope loop at 28.7 K. Each MHL is
labelled by the field in which the sample was initially cooled
down.
}    
\label{Fig. 1}
\end{figure}

\begin{figure} 
\caption{Portions of the MHL initiated from fields lying on the
forward leg (see main panel) and on the reverse leg (see inset
panel) of the envelope loop at 26 K. Each MHL is labelled by the
initial field value.
}
\label{Fig. 2}
\end{figure}

\begin{figure} 
\caption{ MHL loops initiated from FC magnetization values in a
CaRhSn crystal at 1.7 K. For MHL of 2.2 Tesla, we also show the
difference M$_{FC}^{sat}$ - M$^{env}$.
}    
\label{Fig. 3}
\end{figure}

\begin{figure} 
\caption{ Plots of $\Delta$M$_{suc}$ vs field corresponding to MHL
of type-II in (a)YBCO and (b)CaRhSn.
}    
\label{Fig. 4}
\end{figure}
\begin{figure} 
\caption{ The parameter R$_{FC}$ vs field values in which the CaRhSn sample
was field cooled down to 1.7 K.
}    
\label{Fig. 5}
\end{figure}

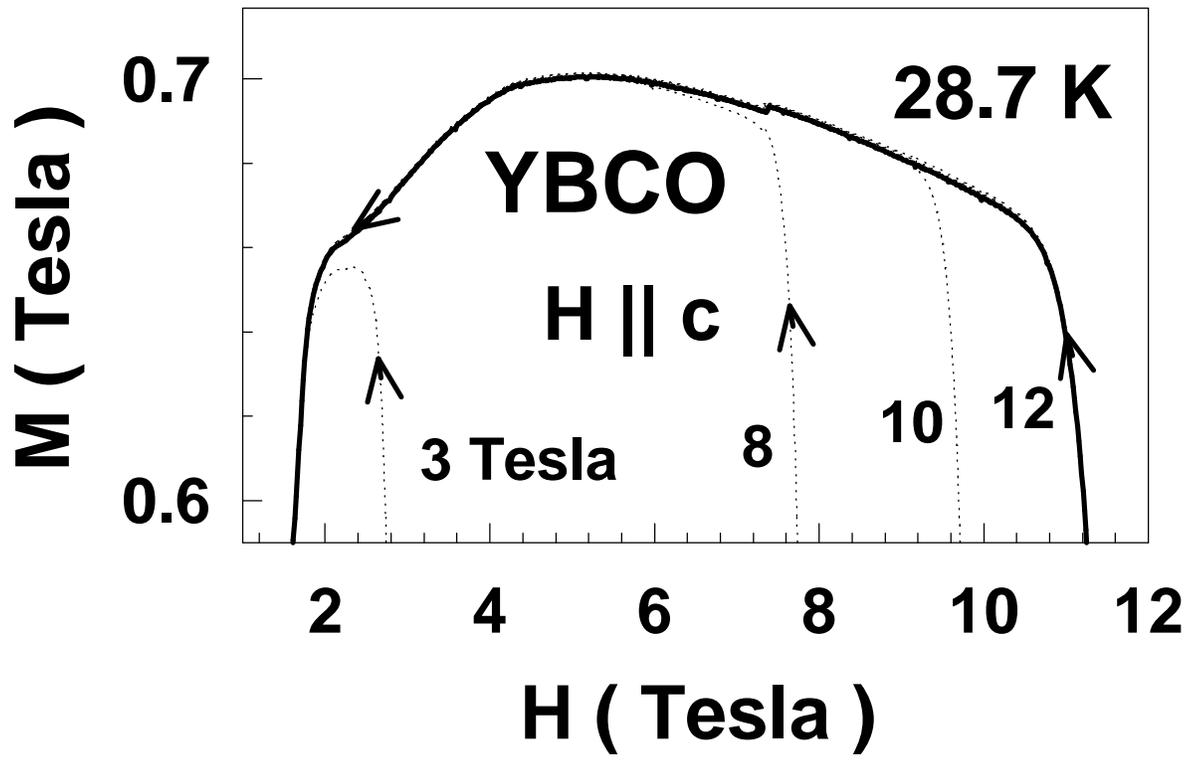
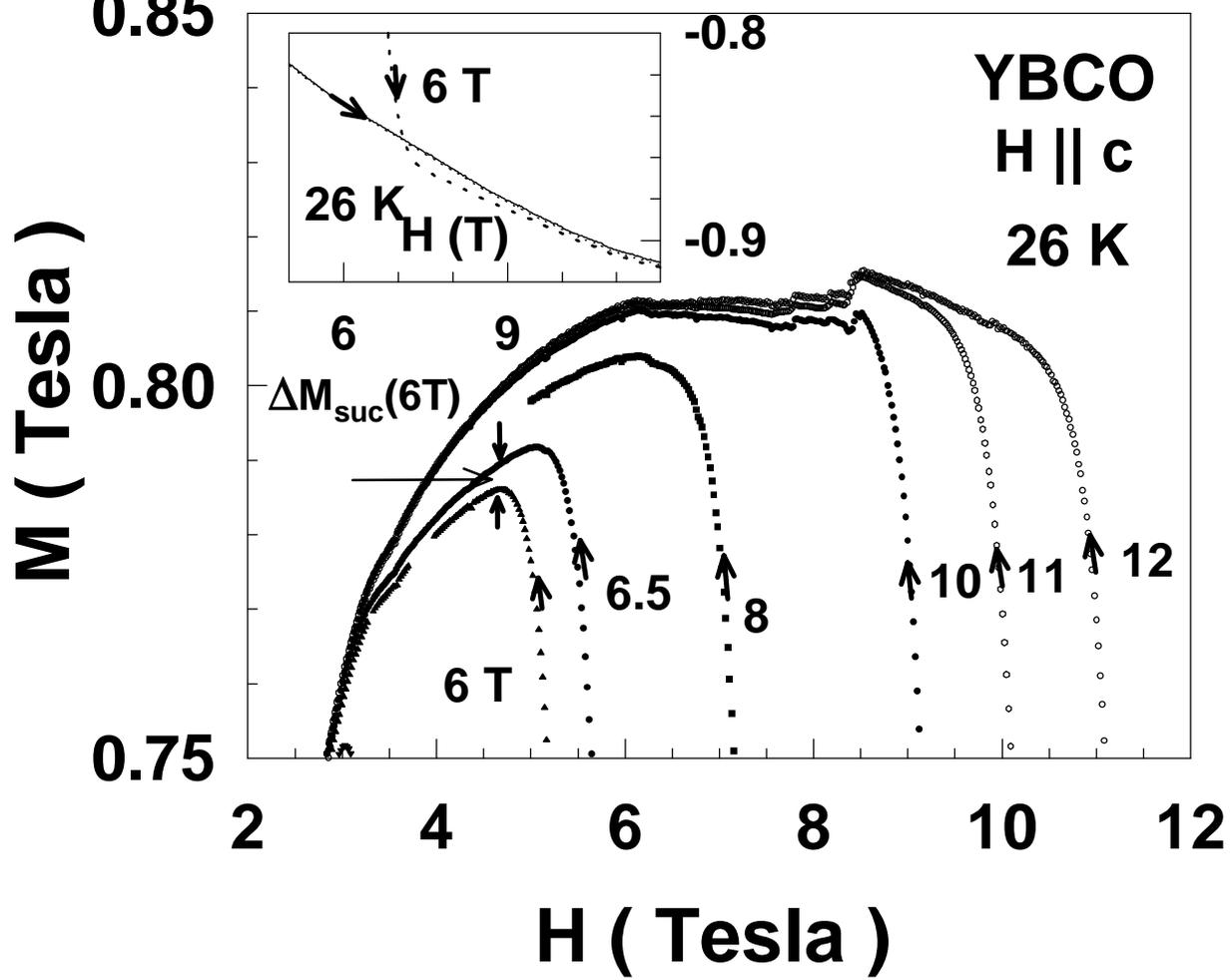
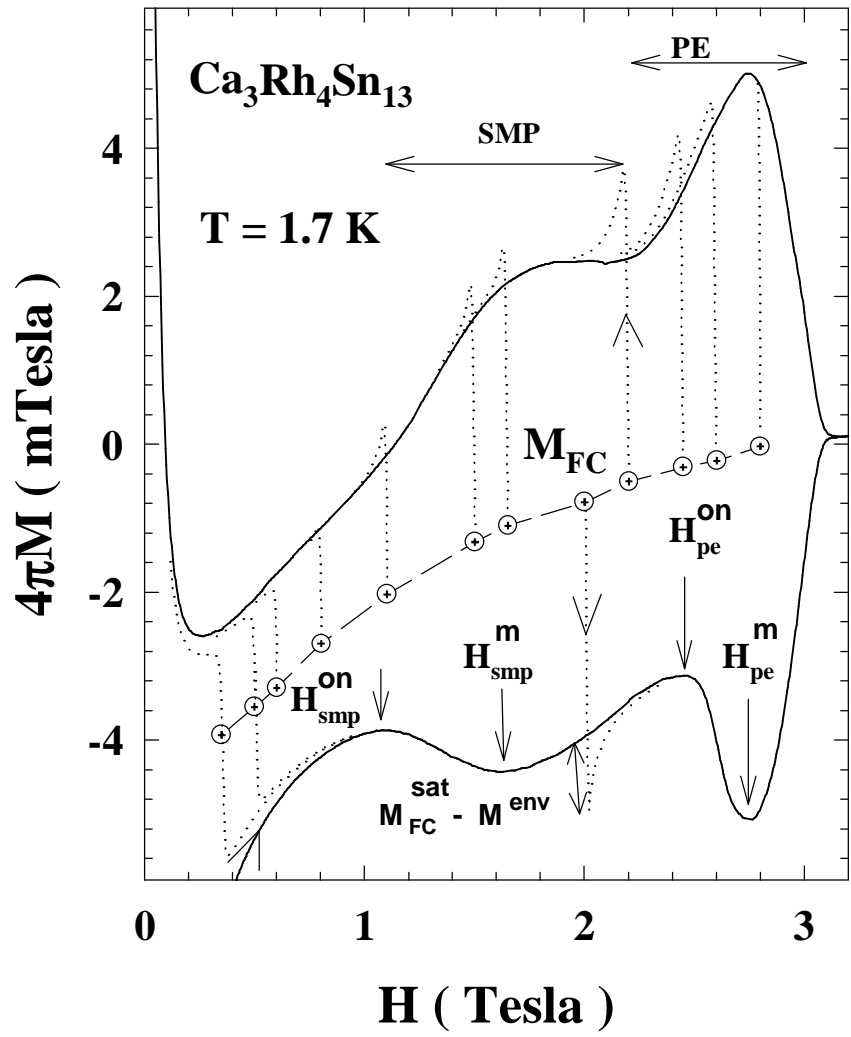
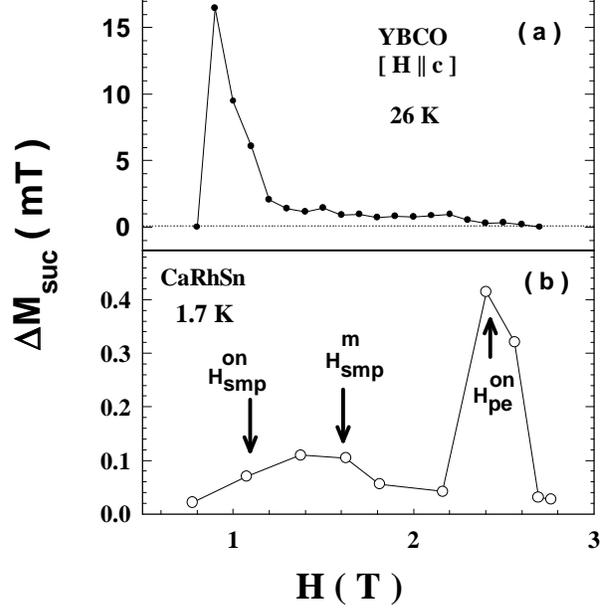
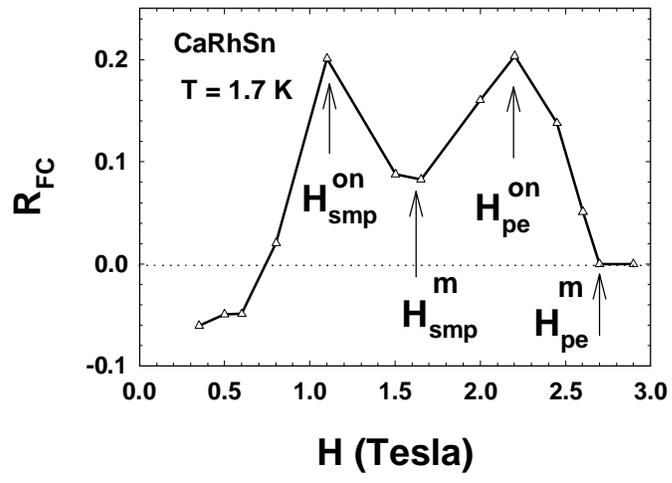
\end{document}